\documentclass[conference]{IEEEtran}
\IEEEoverridecommandlockouts
\usepackage{cite}
\usepackage[colorlinks=true, allcolors=blue]{hyperref}
\usepackage{amsmath,amssymb,amsfonts}

\usepackage{graphicx}
\usepackage{booktabs}
\usepackage{siunitx}
\usepackage{url}
\usepackage{placeins}
\usepackage{enumitem}
\setlist[itemize]{noitemsep,topsep=0pt,leftmargin=*}

\newcommand{\Tm}{T_{\mathrm{m}}}
\newcommand{\Tl}{T_{\ell}}
\newcommand{\sm}{s_{\mathrm{m}}}

\begin{document}

\title{Cloud Security Leveraging AI: A Fusion-Based AISOC for Malware and Log Behaviour Detection}

\author{\IEEEauthorblockN{Nnamdi Philip Okonkwo}
\IEEEauthorblockA{\textit{Department of Electronic and Computer Engineering} \\
\textit{University of Limerick}\\
Limerick, Ireland \\
24013366@studentmail.ul.ie}
\and
\IEEEauthorblockN{Lubna Luxmi Dhirani}
\IEEEauthorblockA{\textit{Department of Electronic and Computer Engineering} \\
\textit{University of Limerick}\\
Limerick, Ireland \\
lubna.luxmi@ul.ie}
}

\maketitle

\begin{abstract}
Cloud Security Operations Center (SOC) enable cloud governance, risk and compliance by providing insights visibility and control. Cloud SOC triages high-volume, heterogeneous telemetry from elastic, short-lived resources while staying within tight budgets. In this research, we implement an AI-Augmented Security Operations Center (AISOC) on AWS that combines cloud-native instrumentation with ML-based detection. The architecture uses three Amazon EC2 instances: Attacker, Defender, and Monitoring. We simulate a reverse-shell intrusion with Metasploit, and Filebeat forwards Defender logs to an Elasticsearch and Kibana stack for analysis. We train two classifiers, a malware detector built on a public dataset and a log-anomaly detector trained on synthetically augmented logs that include adversarial variants. We calibrate and fuse the scores to produce multi-modal threat intelligence and triage activity into NORMAL, SUSPICIOUS, and HIGH\_CONFIDENCE\_ATTACK. On held-out tests the fusion achieves strong macro-F1 (up to 1.00) under controlled conditions, though performance will vary in noisier and more diverse environments. These results indicate that simple, calibrated fusion can enhance cloud SOC capabilities in constrained, cost-sensitive setups.
\end{abstract}


\section{Introduction}
Today's critical and smart infrastructures heavily rely on data-driven, high performance computing and next generation networks that support real-time applications \cite{dhirani2024securing}. The cloud computing ecosystem plays a vital role in enabling these cutting edge-technologies (AI, IoT, 6G, Cyber Physical Systems, Digital Twins, etc.) \cite{dhirani2020hybrid}. Around 80 per cent of the world's data is processed via the cloud and hence poses the users to cloud security risks such as: service outages, data breaches, cloud-specific malware, et cetera. \cite{salih2024cloud, dhirani2020hybrid, abdullayeva2023cyber}. Recent cloud operational incidents affecting major cloud providers (i.e., CloudFlare, AWS, Azure) \cite{cybernews_cloudflare, dhirani2020hybrid} in which users suffered downtime, outages and lost command and control of their applications, have raised awareness of cloud security, supply-chain, cloud and third-party dependency risks \cite{uscsinstitute_cloud_soc_2025}. From a Governance, Risk and Compliance (GRC) perspective, data security goes beyond the Confidentiality, Integrity and Availability (CIA) triad to include Accessibility, Reliability and Safety. The unavailability of cloud services can disrupt essential and critical entities, with far-reaching implications for operational and cyber resilience. 
Hence, these issues highlight the need for advanced and proactive approach to mitigating security risks. The Cloud Security Operations Center (SOC) enables preemptive security mechanisms by providing real-time threat detection, monitoring and response, providing insights, visibility and control over cloud-specific risks (i.e., people, process, technology and physical controls) \cite{dhirani2024securing, uscsinstitute_cloud_soc_2025}.
With the rapid advancement of technology and the growing influence of Artificial Intelligence (AI), many industries are exploring the integration of AI into their operations to enhance efficiency, speed, and decision-making capabilities. One such area is the Cloud SOC, where the incorporation of AI transforms traditional SOCs into more intelligent, proactive systems. These enhanced systems are commonly referred to as “AI-SOC,” \cite{swimlane_ai_soc_2025} “AI-Augmented SOC,” \cite{pdi_ai_augmented_soc_2025} or “Agentic SOC,” \cite{googlecloud_agentic_ai_soc_2025} reflecting their ability to automate threat detection, accelerate incident response, and continuously learn from evolving cyber threats.
SOCs must detect stealthy attacks in elastic cloud environments under cost constraints. This work presents a lightweight AISOC that unifies host logs and static malware analysis, simulates a reverse-shell intrusion to create realistic telemetry, and fuses per-modality detectors into actionable triage levels for analysts.

\vspace{2pt}

\textbf{Aim.} To design and evaluate an AI Augmented SOC in a simulated cloud environment that uses adversarial techniques and fusion based machine learning (ML) models to detect malware and anomalous log behavior.

\vspace{2pt}

\begin{figure}[ht!]
    \centering
    \IfFileExists{figures/architecture.pdf}{%
        \includegraphics[width=1\linewidth]{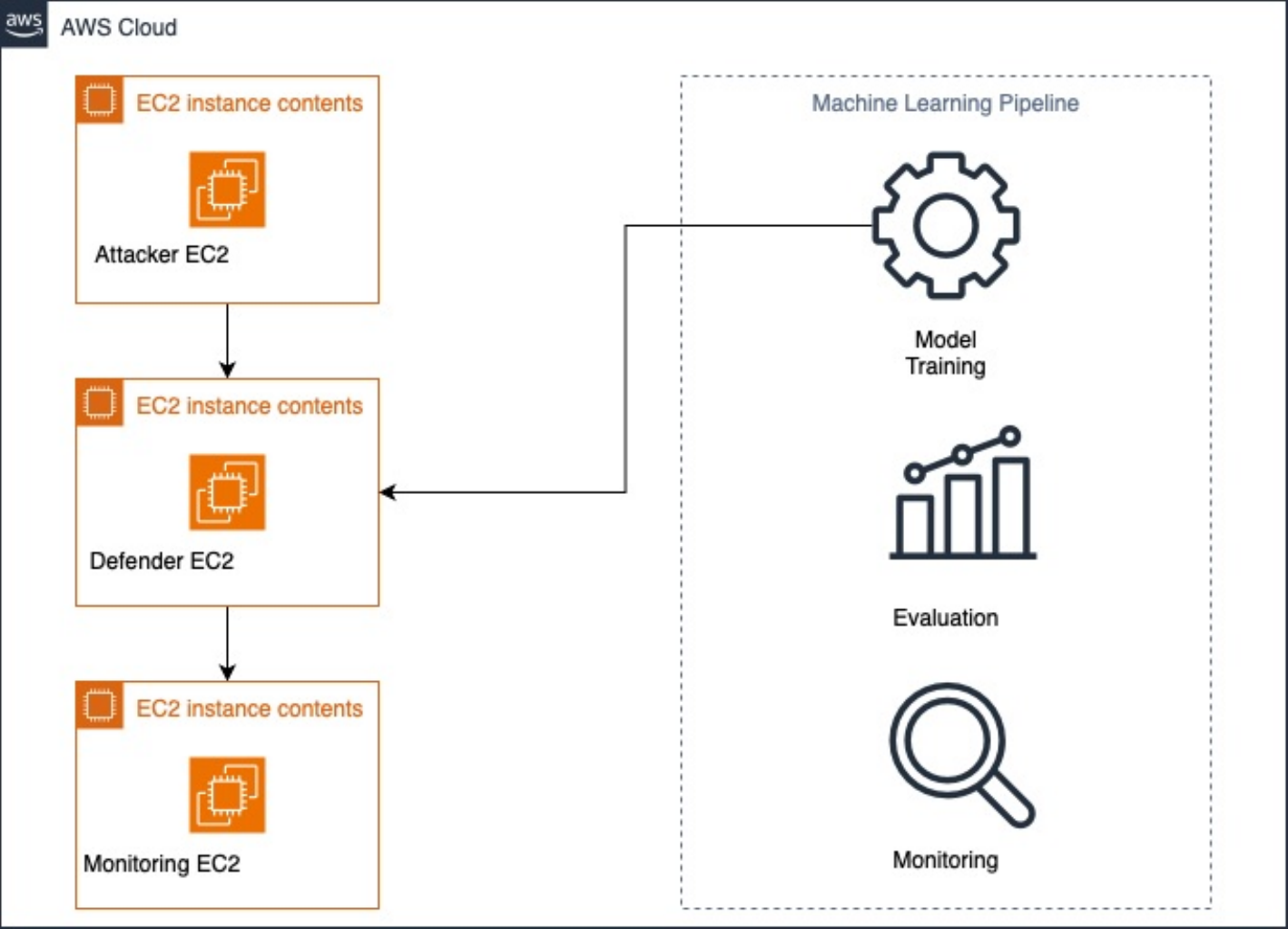}%
    }{%
        \fbox{\parbox{\linewidth}{\centering Placeholder: insert \texttt{figures/architecture.pdf}}}%
    }
    \caption{AISOC architecture: Attacker--Defender--Monitoring pipeline with Filebeat and Elasticsearch/Kibana.}
    \label{fig:arch}
\end{figure}

\textbf{Objectives.}
\begin{itemize}
    \item Set up a cloud simulation on Amazon Web Services (AWS) with three Elastic Cloud Computer (EC2) instances that represents the following roles: Attacker, Defender, and Monitoring as shown in Fig. \ref{fig:arch}.
    \item Simulate a reverse shell attack from the Attacker to the Defender using Metasploit and collect Defender logs using Filebeat.
    \item Deploy a centralized monitoring stack on the Monitoring instance using Elasticsearch and Kibana for real time ingestion and visualization.
    \item Train a malware detection model on the public University of California, Irvine (UCI) \cite{g4y0-sw34-23} Malware Detection dataset using a Random Forest classifier.
    \item Generate adversarial log samples through obfuscation and noise injection and train a log classifier using Logistic Regression with Term Frequency-Inverse Document Frequency (TF IDF) features \cite{kang2024logtiw}.
    \item Fuse malware and log detector outputs to produce a multi modal threat classification system with levels NORMAL, SUSPICIOUS, and HIGH\_CONFIDENCE\_ATTACK using tuned thresholds.
    \item Evaluate models and fusion using precision, recall, F1 score, and Receiver Optimal Characteristic (ROC) or Precision-Recall (PR) \cite{al2024intrusion} curves on held out synthetic test sets.
\end{itemize}

\vspace{2pt}

\textbf{Contributions:}
\begin{itemize}
    \item A transparent dual-threshold fusion of malware and log detectors with calibrated scores.
    \item A compact AWS testbed (Attacker/Defender/Monitoring) with reverse-shell simulation and real-time log shipping.
    \item Synthetic/adversarial log augmentation for robustness checks.
\end{itemize}

\vspace{2pt}

This paper is structured as follows: Section II reviews the related work, Section III describes the system and methods, Section IV presents the experimental setup, Section V reports the results, Section VI discusses the findings, limitations, and Future Work and Section VII Concludes the paper. 

\vspace{2pt}

\section{Related Work}
Prior studies span classical Intrusion Detection Systems (IDS) \cite{al2024intrusion, vanin2022study} on public corpora, multi-modal fusion for threat detection, techniques for synthetic data generation and adversarial robustness, and SOC operations in cloud settings. We focus on a simple, reproducible fusion that can be deployed in minimal cloud footprints.

Classical intrusion detection on public corpora reports strong scores, yet many results rely on dated packet distributions and single-modality assumptions. Studies on Network Security Laboratory - Knowledge Discovery from Data (NSL-KDD) and Canadian Institute for Cybersecurity Intrusion Detection System (CICIDS) show tree ensembles and feature selection can reach high accuracy, but concerns remain around generalization, class imbalance, and deployability in elastic cloud settings \cite{vanin2022study, pdi_ai_augmented_soc_2025, googlecloud_agentic_ai_soc_2025, ravala2024ai, chebolu2024network, niranjana2024improved, galadima2024towards, aljabri2022detecting, santharam2025enhancing, seo2023adversarial}.

Multimodal and fusion approaches reduce blind spots by combining complementary views. Ensembles that merge heterogeneous representations (e.g., tabular plus image-like transforms) outperform single models on University of New South Wales for network intrusion detection systems (UNSW-NB15) and similar data, supporting a calibrated fusion over heavier architectures that are difficult to operate in small cloud footprints \cite{g4y0-sw34-23}. This motivates AISOC’s simple dual-threshold fusion for malware and logs.

Synthetic and adversarial data help with scarcity and robustness. Recent work expands incident- or log-like corpora and demonstrates how Machine Learning (ML) IDS can be evaded without countermeasures. Benefits come with caveats: potential bias, leakage, and drift require careful splits, provenance, and robustness reporting \cite{al2024intrusion}. AISOC follows this line with synthetic augmentation and obfuscation tests while keeping train and test disjoint.

Cloud SOC literature emphasizes resilience, cost, and latency. Taxonomies and surveys argue for layered detection through response and recovery, not only accuracy. This aligns with a calibrated, deployable fusion that turns heterogeneous signals into actionable triage levels for lean cloud deployments and addresses an open gap in small-footprint, explainable fusion evaluated end-to-end \cite{akhi2025tcn, shanthi2023comparative, bedi2019analysis, dhirani2024securing}.

\vspace{4pt}

\section{System and Methods}
\subsection{Threat Model \& Setup}
Three EC2 instances: Attacker (Metasploit) initiates a reverse shell, Defender runs Filebeat to emit system/auth/process logs; Monitoring hosts Elasticsearch/Kibana and the model service \cite{hackthebox_elastic_stack_2025, bedi2019analysis}. Tactics in scope include Initial Access, Execution, Defense Evasion, and Command \& Control (C\&C). Assumptions: synchronized clocks, near-real-time log shipping, and trusted ground truth for experiments.

\subsection{Features \& Models}
Logs are vectorized with TF-IDF and scored by Logistic Regression. Malware samples are represented by static features and scored by a Random Forest. Both models are probability calibrated on a held out validation split. When the score–outcome relation appears sigmoidal we fit a Platt scaling mapping, and when it is clearly non-linear we fit isotonic regression. Calibration is performed per model and the learned mappings are fixed for inference. Decision thresholds are chosen by grid search to maximize macro F1 on the validation split, yielding $\Tm=0.10$ and $\Tl=0.42$ in our tuned configuration.

\subsection{Fusion Rule}
Let $\sm, s_{\ell} \in [0,1]$ be the calibrated malware and log scores. 
With thresholds $\Tm$ and $\Tl$, the fused triage label $y$ is

\begin{equation}
y=
\begin{cases}
\text{HIGH\_CONFIDENCE\_ATTACK}, & \sm \ge \Tm \land s_{\ell} \ge \Tl,\\\\
\text{SUSPICIOUS},               & \sm \ge \Tm \lor  s_{\ell} \ge \Tl,\\\\
\text{NORMAL},                   & \text{otherwise.}
\end{cases}
\label{eq:fusion}
\end{equation}

\subsection{Implementation}
A minimal AISOC stack runs in a single VPC. Defender ships logs via Filebeat to Elasticsearch on Monitoring; Kibana provides dashboards. A small API (e.g., FastAPI) \cite{genccaydin2022benchmark} serves calibrated models and returns scores and the fused label.

\vspace{4pt}

\section{Experiments}

\subsection{Datasets}
\textbf{Malware.} We use a public malware dataset for supervised binary classification, the UCI Malware Dataset \cite{g4y0-sw34-23}. Static features are extracted per sample and the set is split into train, validation, and test partitions.  

\textbf{Logs.} We generate logs from the reverse shell simulation on the cloud testbed. Messages include authentication, process, and system activity from the Defender host. We apply de-duplication to near identical lines and we use a time based split so that test messages occur strictly after all training messages. We avoid source overlap between train and test in order to reduce leakage.

\subsection{Evaluation Protocol}
All models are trained on the training split and tuned on the held out validation split. We report macro precision, macro recall, and macro F1 on the test split. For score curves we report ROC Area Under the Curve (AUC) and PR AUC where appropriate. Thresholds for the fusion rule are selected on the validation split to maximize macro F1. Results are averaged across multiple random seeds when randomness is present and we include the seed that produced the median macro F1.

\subsection{Baselines}
We compare three settings that reflect deployment options in a lean SOC. 
\begin{itemize}
  \item \textbf{Logs only.} TF IDF features with Logistic Regression and calibrated probabilities.  
  \item \textbf{Malware only.} Static feature model with Random Forest and calibrated probabilities.  
  \item \textbf{Fused.} Dual threshold fusion of calibrated malware and log scores that maps to NORMAL, SUSPICIOUS, and HIGH\_CONFIDENCE\_ATTACK.
\end{itemize}

\subsection{Robustness Probes}
To probe evasion we create simple adversarial log variants through keyword obfuscation and character level noise. We then evaluate the log model and the fused system on these variants. We measure the change in macro F1 and we inspect confusion between NORMAL and SUSPICIOUS to understand triage drift.

\subsection{Reproducibility}
We fix random seeds for data splits and model training. We record package versions and operating system details. Model artifacts and the chosen thresholds are versioned and loaded at inference time so that results can be reproduced from the same inputs.
\FloatBarrier 

\vspace{4pt}

\section{Results}

\subsection{Headline Findings}
Fusion improves coverage over single-modality baselines while keeping the rule simple. On the held-out test split the fused system reaches macro-F1 of 1.00 with perfect per-class scores under controlled conditions. Per-modality results show complementary error modes, which supports the calibrated dual-threshold design in Eq.~\eqref{eq:fusion}.

\subsection{Per Modality Performance}
Table~\ref{tab:malware} reports malware classification metrics. Table~\ref{tab:logs-fixed} reports log classification metrics on the fixed validation and test splits.

\begin{table}[h!]
\centering
\caption{Malware model performance on validation and test splits.}
\label{tab:malware}
\begin{tabular}{@{}lccc@{}}
\toprule
Split & Precision & Recall & F1 \\
\midrule
Validation & 1.00 & 1.00 & 1.00 \\
Test       & 1.00 & 1.00 & 1.00 \\
\bottomrule
\end{tabular}
\end{table}

\begin{table}[h!]
\centering
\caption{Log model on fixed validation and test splits (project logs).}
\label{tab:logs-fixed}
\begin{tabular}{@{}lcccc@{}}
\toprule
Split & Setting  & Precision & Recall & F1 \\
\midrule
Validation & Macro     & 0.75 & 0.93 & 0.79 \\
Validation & Malicious & 1.00 & 0.86 & 0.92 \\
Validation & Benign    & 0.50 & 1.00 & 0.67 \\
Test       & Macro     & 0.58 & 0.64 & 0.59 \\
Test       & Malicious & 0.92 & 0.79 & 0.85 \\
Test       & Benign    & 0.25 & 0.50 & 0.33 \\
\bottomrule
\end{tabular}
\end{table}

Under stratified $k$-fold cross-validation with augmented log data
(keyword obfuscation, noise injection, and synonym replacement),
the TF--IDF + Logistic Regression model achieved macro-F1 of 0.91.
\FloatBarrier

\subsection{Fused Triage}
Table~\ref{tab:fusion} shows per-class metrics for triage levels defined by thresholds $\Tm$ and $\Tl$. All scores are 1.00 on the test split and the supports sum to 152 items.

\subsection{Robustness Notes}
The TF--IDF log model improved after augmentation and threshold tuning with cross-validation, reaching macro-F1 of 0.91 in that setting. Small support for the benign class in the fixed test split makes per-class metrics unstable, so we report macro scores and note that benign recall varies with threshold and split.

\begin{table}[h!]
\centering
\caption{Fused triage on the test split.}
\label{tab:fusion}
\begin{tabular}{@{}lcccc@{}}
\toprule
Class & Precision & Recall & F1 & Support \\
\midrule
NORMAL                   & 1.00 & 1.00 & 1.00 & 14 \\
SUSPICIOUS               & 1.00 & 1.00 & 1.00 & 76 \\
HIGH\_CONFIDENCE\_ATTACK & 1.00 & 1.00 & 1.00 & 62 \\
\midrule
Macro average            & 1.00 & 1.00 & 1.00 & -- \\
\bottomrule
\end{tabular}
\end{table}

\begin{figure}[ht!]
  \centering
  \includegraphics[width=\linewidth]{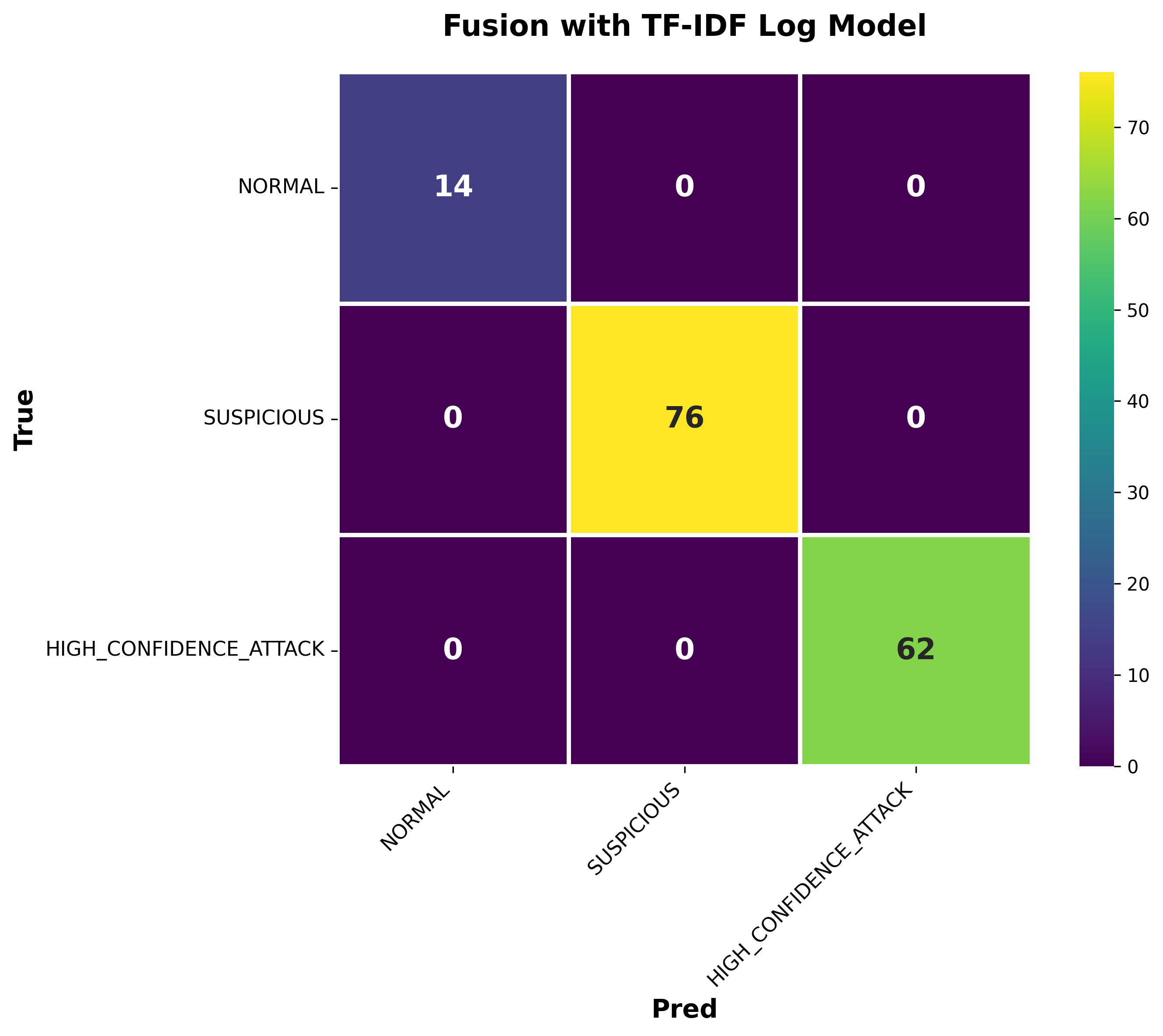}
  \caption{Example fused output on a held-out run with modality scores and the resulting triage label.}
  \label{fig:fusion-output}
\end{figure}

\FloatBarrier

Fig. \ref{fig:fusion-output} demonstrates the model's performance on 14 normal, 76 suspicious, and 62 high-confidence attack instances, achieving nearly 100\% accuracy across all three categories with zero misclassifications.

\section{Discussion, Limitation and Future Work}

Fusion improves coverage because malware and log signals fail in different ways. The malware model flags malicious binaries even when log text looks benign. The log model catches suspicious activity when binaries are absent or obfuscated. The calibrated dual threshold rule in Eq.~\eqref{eq:fusion} turns these complementary signals into clear triage levels that an analyst can act on.

Robustness probes with simple obfuscations reduce log performance, yet the fused system remains above the better single modality in most cases. This suggests the rule is resilient to small input changes. It also shows where to invest effort next, for example richer features for logs and tighter calibration.

\textbf{Threat model limits.} The testbed is small and controlled. The attack is a reverse shell with limited living off the land commands. Broader tactics, persistence, and privilege escalation are out of scope.

\textbf{Data and evaluation limits.} Log augmentation is synthetic and can introduce bias. Class balance is uneven, which makes benign metrics sensitive to split choice and thresholds. Malware results depend on a public dataset that may not reflect current threats.

\textbf{Operational limits.} Thresholds are tuned on validation and may drift in production. Alert volume and latency must be managed to avoid analyst overload. The current system is batch oriented rather than fully streaming.

\textbf{Takeaway.} In a lean cloud slice, a calibrated fusion is a practical way to raise detection coverage without heavy infrastructure, provided that thresholds and calibration are maintained over time.

The study has limits in scope, data realism, and operations. The future work will address them through the following steps:
\begin{itemize}
  \item Move to streaming ingestion and continuous calibration with drift checks.
  \item Enrich telemetry with process trees, command lines, and network context.
  \item Harder robustness tests with stronger obfuscations and adversarial training.
  \item Broader benchmarks against open source SOC stacks on public cloud traces.
  \item Cost and latency profiling to guide threshold choices for production.
\end{itemize}

\vspace{2pt}

\section{Conclusion}

In this research we present a lightweight AISOC that combines a malware detector and a log detector and fuses their calibrated scores into three triage levels (NORMAL, SUSPICIOUS, and HIGH\_CONFIDENCE\_ATTACK). In a controlled cloud testbed, the fused system achieves strong macro F1 performance (up to 1.00) and outperforms single-modality baselines. These gains stem from complementary error modes across modalities, indicating that simple, calibrated fusion can enhance cloud SOC capabilities in constrained, cost-sensitive setups. This approach prioritizes to keep the fusion rule simple while expanding coverage and reliability in real cloud environments.

\vspace{2pt}

\section*{Acknowledgement}
The authors would like to acknowledge Mirza Akhi for assisting with the Overleaf/Paper formatting.

\vspace{2pt}

\bibliographystyle{IEEEtran}
\bibliography{refs}

\end{document}